\documentclass[12pt]{article}

\usepackage{mheck}

\institution{HARVARD}{\   Jefferson Physical Laboratory, Harvard University, Cambridge, MA 02138, USA}
\institution{HarishChandra}{\ Harish-Chandra Research Institute, Chhatnag Road, Jhusi, Allahabad 211019, India }

\title{ About the Absence of Exotics \\ and the Coulomb Branch Formula}
\authors{Michele Del Zotto \worksat{\HARVARD}\footnote{e-mail: {\tt eledelz@gmail.com}} and Ashoke Sen \worksat{\HarishChandra}\footnote{e-mail: {\tt sen@hri.res.in}}}

\abstract{ The absence of exotics is a conjectural property of the spectrum of BPS states of four--dimensional $\cn=2$ supersymmetric QFT's. In this letter we revisit the precise statement of this conjecture, and develop a general strategy that, if applicable, entails the absence of exotic BPS states. Our method is based on the Coulomb branch formula and on quiver mutations. In particular, we obtain the absence of exotic BPS states for all pure SYM theories with simple simply--laced gauge group $G$, and, as a corollary, of infinitely many other lagrangian $\cn=2$ theories.}

\begin{document}

\maketitle

\tableofcontents

\section{Introduction}

The study of the BPS spectrum of $\cn=2$ supersymmetric gauge theories has been an active area
of research in recent times. One of the conjectured properties of this spectrum
is the absence of exotics\cite{GMN_framed}. The latter is the statement that, factoring out the contribution of the Clifford vacuum, any BPS state transforms with respect to the rotational SU(2) and the SU(2) R-symmetry as a singlet of R-symmetry and an arbitrary (reducible) representation of the rotational symmetry.

It turns out that this property of the BPS spectrum is not a protected information: the spectrum
can change from non-exotic to exotic as we vary the parameters of the theory. This makes a direct proof of the conjecture very
difficult. However, one can try to prove a related statement
for the class of theories whose BPS spectrum has a quiver description in some region of the
moduli space. In the region where the quiver description is valid, 
the absence of exotics can be related to the statement that the
Hodge numbers $h^{m,n}$ of the corresponding quiver moduli space are non-zero only
for $m=n$\cite{DiacoMoore,FelixKlein,MMgeom,CN}. Since the Hodge diamond is a topological information, the vanishing of $h^{m,n}$ for $m\ne n$
is protected under small deformations and hence this should be easier to prove. This will
establish that the spectrum of the corresponding gauge theory is exotics-free in the
region of parameter space where the quiver description is valid.

If we introduce a generating function $Q(y,t)$ of the Hodge numbers $h^{m,n}$ with $y$ (resp. $t$)
being the variable conjugate to $m+n$ (resp. $m-n$), the absence of exotics, as defined above, corresponds to the $t$ independence of
$Q$. $Q(y,t)$ can jump across walls of marginal stability but the $t$ independence of $Q$
must hold in all chambers.
In this paper we prove this for quivers associated with pure 
$\cn=2$ supersymmetric gauge theories with simple simply-laced gauge groups. The main ingredient of the
proof is the Coulomb branch formula\cite{MPS1,MPS2,EquivalentWC,MPS3,MPS4,MPS5,MPSreview} that expresses $Q(y,t)$ 
in terms of a set of unknown functions 
$\Omega_S$ of
$t$ called the single centered indices. This formula was derived by demanding that $Q(y,t)$
satisfies the appropriate wall crossing formula. We show that for the quivers associated with 
pure gauge theories the $\Omega_S$'s which could introduce $t$ dependence of $Q$ vanish
identically by the requirement of mutation symmetry. This not only proves the $t$
independence of $Q$, but also gives us the stronger result that the Coulomb branch
formula can determine the Hodge numbers of these quivers in all chambers without any
further input in the form of the single centered indices. Moreover, having showed this fact for the category of BPS states of these models, the absence of exotics extends to all subcategories.

Recently, with completely different methods, the vanishing of $h^{m,n}$ for $m\ne n$ was shown to hold for $SU(k)$ SYM \cite{MMgeom}. The same method should extend (in principle) to all 4d $\cn=2$ UV-complete models obtained from the geometric engineering of the Type IIA superstring theory on toric CY--three folds. We have not yet been able to determine necessary conditions for the applicability of our strategy to models with BPS quivers. It would be very interesting to understand whether the two classes agree or not.\footnote{ As remarked by G. W. Moore during the JPM Lyon 2014.}

The rest of the paper is organized as follows. In \S\ref{sstate} we outline the problems which
arise if we try to directly prove the absence of exotics conjecture in
$\cn=2$ supersymmetric  gauge theories, and describe how
we can make this into a well defined question by posing the problem in terms of the associated
quiver.  In \S\ref{sstrategy} we outline the strategy that we use for proving the vanishing of the single
centered indices for a class of quivers. In \S\ref{SYM} we apply this to prove the vanishing
of the single centered indices for the quivers associated with the pure $\cn=2$ supersymmetric
gauge theories with simply laced gauge groups and in \S\ref{sspecial} we generalize this analysis to
some other related theories. Appendix \ref{sa} contains a specific example of how the BPS
spectrum of a gauge theory can change from non-exotic to exotic under continuous deformations
and appendix \ref{ovvieta} reviews a different proof of the absence of exotics in a different class
of theories, the models of class $\cs[A_1]$ with a BPS quiver description. For this class of systems, indeed, the absence of exotics is a trivial corollary of the recent Geiss--Labardini-Fragoso--Schr\"oer theorem \cite{superpowers}.
In appendix \ref{omegazz} we show how for a subset of theories described in 
\S\ref{sspecial} we
can directly prove the vanishing of the single centered indices using our method.

\section{Statement of the problem} \label{sstate}
Consider a four dimensional $\cn=2$ supersymmetric gauge theory with gauge group $G$ of rank $r$. The model might also carry a global flavor symmetry group $F$ of rank $f$. Let it flow on its Coulomb branch by giving vevs to the scalars in the $\cn=2$ vectormultiplets thus breaking $G$ to its Cartan 
subgroup
$U(1)^r$. In addition,  also the flavor symmetry might be broken to $U(1)^f$ by a choice of generic mass deformations. The low energy internal quantum numbers of the excitations of this model in this regime are $r$ electric, $r$ magnetic, and $f$ flavor charges that are conserved and quantized and thus are valued in a rank $k \equiv 2r + f$ integer lattice $\Gamma$, the charge lattice of the model. Let us denote by $\mathscr{P}$ the space of all parameters of the model in this phase (UV gauge couplings, Coulomb branch parameters, masses,...). The central charge of the $\cn=2$ superalgebra defines a map $Z \colon \mathscr{P} \rightarrow \text{Hom}(\Gamma,\C)$. Supersymmetric representation theory entails that for a given point $p\in\mathscr{P}$ all the excitations of the model with charge $\gamma\in\Gamma$ have masses that obey the BPS bound: $M(p;\,\gamma)\geq |Z(p;\,\gamma)|$; the BPS excitations are those that saturate this bound. In particular, BPS excitations always come in short $\cn=2$ multiplets. With respect to the bosonic $SU(2)_{\text{spin}}\otimes SU(2)_{R}$ symmetries of the 4D $\cn=2$ superalgebra, short multiplets have Clifford vacuum in the representation
\be \label{eone}
\rho_{hh} \, : \, (1/2, 0) + (0, 1/2)\, .
\ee
In contrast the Clifford vacuum of a long multiplet has representations
\be \label{etwo}
\rho_{hh} \otimes \rho_{hh}\, :(1,0) + (0,0) + (0,1) + (0,0) + (1/2, 1/2) + (1/2, 1/2)\, .
\ee
Above, we have used the notation in which $(j, j^\prime)$ will denote a multiplet carrying $SU(2)_{spin}$
angular momentum
$j$ and $SU(2)_R$ angular momentum $j^\prime$. 
The structure of a multiplet is determined by the choice of a representation $\mathfrak{h}$ of $SU(2)_{\text{spin}}\otimes SU(2)_{R}$ to be tensored with the Clifford vacuum. 
A strong version of the no-exotics conjecture says that all the short multiplets in the spectrum are obtained by taking the
tensor product of \eqref{eone} with $SU(2)_{\text{spin}}\otimes SU(2)_{R}$ representations of the
form $\oplus_i (j_i,0)$.

\subsection{Difficulties with the no--exotics conjecture}

We now elaborate on possible difficulties in establishing this conjecture in the strong form above. Clearly
masses of non--BPS excitations also depend on the parameters of the theory. 
Let $\psi$ denote a non--BPS state of the system, with charge $\gamma\in\Gamma$. 
Now assume that there is a codimension one locus in the parameter space 
where $\psi$ accidentally satisfies the BPS--bound, captured by the equation
\be\label{3wall}
M(p;\psi) = |Z(p,\gamma)| \qquad p\in\mathscr{P}.
\ee
This is consistent with \textsc{susy} representation theory: since $\psi$ comes in a long multiplet $\rho_{hh}\otimes \rho_{hh}\otimes \mathfrak{h}_\psi$, the corresponding spurious BPS--state can be thought as part of a (reducible) short multiplet $\rho_{hh}\otimes \mathfrak{h}^\prime$ with $\mathfrak{h}^\prime \equiv \rho_{hh}\otimes \mathfrak{h}_\psi$. To be consistent with the literature about the wall--crossing phenomenon, we would like to refer to the codimension one loci defined in eqn.\eqref{3wall} as  walls of the third kind. Clearly, on this wall it is very hard to distinguish the original BPS--states from the spurious ones described above. 
The precise structure of these walls are determined by conditions about the non-BPS spectrum, and these conditions are typically not protected by \textsc{susy}. For this reason walls of the third kind are very hard to control.

Walls of the third kind contains exotic states since $\rho_{hh}\otimes \mathfrak{h}^\prime$ has
states in non-trivial $SU(2)_R$ representations.\footnote{ There may
be further complications due to the mixing of single particle states with multi particle
states involving massless states in the gauge multiplet, but we shall not discuss
this here. However, these subtleties do not affect the index: two non--identical particles are described by the tensor product of the corresponding Hilbert spaces, and the tensor product of more than one short multiplet is in a long multiplet.}
However, the worse aspect of this phenomenon is that as we cross this wall, a different set of
short multiplets may combine into a long multiplet and acquire mass above the BPS bound,
leaving behind 
a spectrum of short multiplets that differs from the original one.
In particular, even if the original spectrum of BPS states was free from exotic states, the spectrum 
obtained after crossing the wall can contain exotic states. An example of this has been
given in appendix \ref{sa}.

One might wonder whether the above mentioned problem could be avoided by 
studying an appropriate supersymmetric index. Indeed 
at any given point of parameter space one can define an index that vanishes on long multiplets by construction, and counts the BPS states in a given charge--superselected sector of the one--particle Hilbert space\cite{GMN_framed}.\footnote{ For a nice review see \cite{FelixKlein}.} Let us denote by $\ch_{\text{BPS},\,p}(\gamma)$ the subspace of the one--particle Hilbert space consisting of particles with charge $\gamma\in\Gamma$ that satisfy the BPS--bound at $p\in\mathscr{P}$. Let $J_3$ and $I_3$ be two generators of $SU(2)_\text{spin}$ and $SU(2)_R$ respectively. The index in question is \cite{GMN_framed}
\be
\text{Tr}_{\,\ch_{\text{BPS},\,p}(\gamma)} \, \left[(2 J_3) (-1)^{2 J_3} (-y)^{2(I_3 + J_3)}\right].
\ee
Clearly, the Clifford vacuum of short multiplets can always be factored out from the index, defining $\Omega(p, \gamma; y)$, the protected spin character \cite{GMN_framed}:
\be \label{edefomega}
 \text{Tr}_{\,\ch_{\text{BPS},\,p}(\gamma)} \, \left[(2 J_3) (-1)^{2 J_3} (-y)^{2(I_3 + J_3)}\right]
  = (y - y^{-1})\,\Omega(p, \gamma; y).
\ee
This index does not change across walls of the third kind since this involves appearance
and disappearance of long multiplets. However this index does not carry information 
about whether a contribution comes from exotic states or non-exotic states. In particular
one can easily verify that the spectrum of short multiplets on two sides of the wall in the
example of appendix \ref{sa} has exactly the same index, even though one has no exotics
and the other one has exotic states.

This phenomenon makes it difficult to give a completely convincing proof of the
absence of exotics in gauge theories 
since the beheaviour of non-BPS states is hard to study, particularly at strong coupling.
For this reason we shall address a slightly different but related question. This will be
explained in \S\ref{squiver}.


\subsection{No exotics conjecture for BPS quivers} \label{squiver}



Assume that the charge lattice $\Gamma$ of the system has the quiver property \cite{CV11, BPSQ}. This means that there is a basis $\{e_1,...,e_k\}$ of $\Gamma$ such that ({\it i }) at a point $p\in\mathscr{P}$ there is a $\theta\in[0,2\pi)$ such that $\text{Im}(e^{-i\theta} Z(e_i,p))>0$ for all $i=1,...,k$; ({\it ii }) all $e_i$'s are charges of hypers that are stable at $p\in\mathscr{P}$, and ({\it iii }) if $\gamma = \sum_{i=1}^k n_i e_i$ is the charge of a BPS excitation, then either all $n_i$'s are positive integers, or all $n_i$'s are negative \cite{CV11, BPSQ}. Any choice of basis with the quiver property comes with a somewhat artificial splitting of the charge lattice into a cone of particles (all $n_i$ positive) and a cone of anti--particles (all $n_i$ negative): $\Gamma \simeq \Gamma^+ \cup (-\Gamma^+)$. By \textsc{pct} symmetry, all these choices are equivalent. 
If $\Gamma$ has the quiver property, at points $p\in\mathscr{P}$ such that all the $Z(e_i,p)$'s are nearly aligned, BPS particles can be described as atom--like objects, bound--states of elementary constituents kept together by the $r$ abelian interactions that governs the dynamics in the Coulomb phase \cite{Denef}. 
In this regime, a convenient description of the IR dynamics of a single BPS particle of charge $\gamma=\sum_{i=1}^k n_i e_i\in\Gamma^+$ is obtained using the supersymmetric quiver quantum mechanics (SQM) with four supercharges that describes its worldline. The gauge group of such SQM is $U(n_1)\times U(n_2) \times \dots \times U(n_k)$, and the matter content is determined as follows: let us denote with $\langle \, , \, \rangle_D$ the Dirac-Schwinger-Zwanziger sympletic
product between the elements of $\Gamma$, whenever $\langle e_i \, , e_j \rangle_D \equiv B_{ij} > 0$, there are $B_{ij}$ bifundamental chiral multipliets charged under $U(n_i)\times U(n_j)$ transforming in the $(\mathbf{\bar{n}}_i,\mathbf{n}_j)$. In addition, whenever one can form a gauge invariant single trace operator out of these bifundamentals, this can give rise to a contribution to the SQM superpotential.\footnote{ Notice that here we are assuming that at very low energy all fields giving rise to quadratic contributions in the superpotential can always be safely integrated out. This assumption is equivalent to the requirement that the superpotential is generic ``enough'', see section 7 of \cite{DWZ1}.} Whether the given charge $\gamma$ corresponds to a stable BPS particle or not depends on the presence or absence of \textsc{susy} ground states for the SQM. This, in turn, is determined by the $F$-term
and $D$--term constraints \cite{Denef,CV11,BPSQ}. Let us denote by $\cm(\gamma,p)$ the moduli space of \textsc{susy} vacua of the SQM associated with a BPS particle of charge $\gamma$ at the point $p\in\mathscr{P}$. The central charge of the 4d $\cn=2$ superalgebra determines the FI terms entering the $D$--term constraints for the SQM description of the BPS particle as follows:
\be
\zeta_i(\gamma,p) = \text{Im} \left( Z(e_i,p) / Z(\gamma,p)\right)\, .
\ee
Depending on their value at $p\in\mathscr{P}$, $\cm(\gamma,p)$ might be empty or not. If $\cm(\gamma,p)$ is empty, there are no \textsc{susy} vacua corresponding to the charge in question at the given point of the parameter space and the BPS particle is unstable. If, instead, $\cm(\gamma,p)$ is non--empty, the BPS particle is stable and its $SU(2)_\text{spin}\otimes SU(2)_R$ quantum numbers are determined by quantizing it. 

Let us first assume that $\gamma$ is primitive, \emph{i.e.} that $\text{gcd}(n_1,n_2,\dots,n_k)=1$: in this case it is believed that the moduli spaces $\cm(\gamma,p)$ are compact, projective, and smooth complex algebraic varieties. 
If $\mathfrak{h}$ denotes the representation with which \eqref{eone} is tensored to give
the spin and R-isospin content of the given BPS state of charge $\gamma$, then
the $\mathfrak{h}$ representation can be read off from the Hodge diamond of $\cm(\gamma,p)$: $J_3$ is encoded in the $SU(2)$ Lefschetz action on the Dolbeault cohomology \cite{Witten,Denef,BPSQ}, while $I_3$ corresponds to a ``Hodge'' $SU(2)$ \cite{DiacoMoore,MMgeom}.\footnote{ More concretely: $2 J_3 = m+n - d$ where $d\equiv\dim_\C \cm(\gamma)$, and $2 I_3 = m - n$ on $\bigoplus_{m,n} H^{m,n}(\cm(\gamma,p))$. Notice that the existence of this $I_3$ is predicted by \textsc{susy} and, as such, it is a (very stringent) necessary condition to interpret a given quiver as a BPS quiver: if the off--diagonal Dolbeault cohomology of the moduli space of a stable representation of a given quiver is not an $SU(2)_R$ representation, the given quiver cannot be a BPS quiver.} In particular 
\cite{MMgeom},
\be\label{poincarep}
Q(\gamma, p; y,t) \equiv \text{Tr}_{\mathfrak{h}} \left[ (-y)^{2J_3} t^{2I_3}\right]
= \sum_{m,n\in\mathbb{Z}}(-y)^{m+n-d}\, t^{m-n} \, h^{m,n}(\cm(\gamma,p))\,
\ee
where $h^{m,n}$ are the Hodge numbers of $\cm(\gamma,p)$.
The absence of exotics in the spectrum translates to the
statement that $Q(\gamma, p; y,t)$ is independent of $t$, i.e.
\be \label{enoexotics}
h^{m,n}=0 \quad 
\text{for} \quad m\ne n\, .
\ee

Note that the quiver description
of BPS states was obtained by working in a particularly simple
regime of the parameter space of the gauge theory. In particular it
does not know anything about non-BPS states and possible existence of
the walls of the third kind: even if the gauge theory moduli pass through a
wall of the third kind and the spectrum of BPS states change, the quiver spectrum remains unchanged across such walls. Thus, on the other side of a wall of the third kind the 
spectrum of BPS states of the quiver will not agree with that of gauge theory. 
On this side the BPS quiver captures only the protected \textsc{susy} indices we discussed in the previous section.
Nevertheless we shall take \eqref{enoexotics} as the definition of the
no exotics conjecture and try to prove this for a class of quivers.

If $\gamma$ is not primitive the moduli spaces $\cm(p,\gamma)$ are not expected to be compact anymore. In this case $Q(\gamma, p; y,t)$ has a more refined definition \cite{KS1,KS2,Joyce4}, that correctly quantize $\cm(\gamma,p)$. With the correct quantization, the absence of exotics is stated as above ($t$ independence of $Q(\gamma, p; y,t)$ at all 
$p\in\mathscr{P}$ for all $\gamma$).

We end this discussion by writing down the expression for the protected spin
character $\Omega(\gamma,p;y)$
defined in \eqref{edefomega}.
First we compute $\text{Tr}_{\text{BPS}} \left[ (-y)^{2J_3} t^{2I_3}\right]$
by taking the product of \eqref{poincarep} with the contribution from the
representation \eqref{eone}:
\be 
\text{Tr}_{\,\ch_{\text{BPS},\,p}(\gamma)} \left[ (-y)^{2J_3} t^{2I_3}\right]
= (-y-y^{-1} + t + t^{-1}) Q(\gamma, p; y,t)\, .
\ee
This gives
\begin{eqnarray}
\text{Tr}_{\,\ch_{\text{BPS},\,p}(\gamma)} \, \left[(2 J_3) (-1)^{2 J_3} (-y)^{2(I_3 + J_3)}\right]
&=& \left[v{\partial\over \partial v} 
\text{Tr}_{\,\ch_{\text{BPS},\,p}(\gamma)} \left[ (-v)^{2J_3} t^{2I_3}\right]\right]_{t=-y,v=-y}
\nonumber \\
&=& (y - y^{-1}) Q(\gamma;p;-y,-y)\, .
\end{eqnarray}
Using \eqref{edefomega} and \eqref{poincarep} we now get
\be
\Omega(\gamma,p;y) = \sum_{m,n\in\mathbb{Z}} (-1)^{m-n} y^{2m-d} h^{m,n}(\cm(\gamma,p)).
\ee
Let us conclude this section with the following {\it caveat}: while the definition of the index $\Omega$ is not
affected by possible mixing in between single and multi-particle states in the gauge theory, 
the definition of $Q$ itself could be affected. However, as long as we consider generic points of the moduli space, 
the only massless particles in the spectrum are precisely the abelian vectormultiplets on the Coulomb branch and  
the contribution of these can be safely factored out giving rise to an effective definition of a single-particle Hilbert space.
This is the space that enter in the definition of $Q$ above. Away from these generic points we cannot ignore such effects anymore and, again, the BPS quiver description of the spectrum is no longer valid, even if the indices all agree with those computed using the BPS quivers.

\section{The strategy} \label{sstrategy}
We shall first review some pertinent aspects of the Coulomb branch formula\cite{MPSreview} and
then outline the general strategy for proving $t$ independence of 
$Q(\gamma, p; y,t)$.

\subsection{Review of Coulomb Branch Formula}

We shall begin by reviewing the basic structure of the Coulomb branch formula for
$Q(\gamma;p;y,t)$. It takes the form
\be \label{eqexp}
Q(\gamma;p;y,t) = \sum_{n\ge 1} \sum_{ \alpha_i\in \Gamma^+,m_i\in \Z, m_i\ge 1
\atop {\sum_{i=1}^n m_i \alpha_i=\gamma }
}
{\cal F}(\alpha_1,\cdots \alpha_n; m_1,\cdots m_n; 
\zeta(\gamma, p); y) \prod_{i=1}^n \Omega_S(\alpha_i;y^{m_i}, t^{m_i})\, ,
\ee
where ${\cal F}(\alpha_1,\cdots \alpha_n; m_1,\cdots m_n;\zeta(\gamma, p); y)$ is a function of its arguments.
The algorithm for computing ${\cal F}$ can be found in \cite{MPSreview} but we shall not need the
details. The important point for us is that ${\cal F}$ 
does not depend on $t$. 
The only dependence on $t$ in the Coulomb branch formula is introduced by the functions 
$\Omega_S(\a\,;y,t)$, called single centered indices. 
Some crucial properties of the single centered indices are the following:
\begin{eqnarray}\label{crucial2}
&\Omega_S(e_i\,;y,t) \equiv 1 \quad \text{and} \quad \Omega_S(\ell e_i\,;y,t) \equiv 0 \quad \forall \, i,
\quad  \forall \, \ell\ne 1
\\\label{crucial}
&\Omega_S(\gamma\,;y,t) \equiv 0 \quad\forall \,\gamma \notin \Gamma^+\, .
\end{eqnarray}
Moreover, $\Omega_S$ vanishes on all $\gamma$ that have support on disjoint subquivers. 
For generic $\gamma\in\Gamma^+$, and for generic superpotentials, $\Omega_S$ are conjectured to be independent of $y$ and universal unknown functions of $t$. 
We shall in fact not need to assume the $y$ independence of $\Omega_S$ since we shall
show that for the class of quivers we shall analyze,  $\Omega_S$ actually vanishes
for any $\gamma\in\Gamma^+$
that is not a basis vector. In the context of quivers, 
$\Omega_S$ first appeared in  \cite{1205.5023,1205.6511,1207.0821} 
where they were computed for cyclic quivers
with rank 1 at each node, and refered to as the degeneracies of 
pure Higgs states or intrinsic Higgs
states.

Another very important aspect for the purpose of this note is the behavior of the single center indices with respect to quiver mutations. As we have stressed in \S\ref{squiver}, different choices of basis of the charge lattice correspond to different splitting of it in between cones of particles and anti--particles. \textsc{pct} symmetry implies all these choices are equivalent. However, to each different (but equivalent) choice of basis correspond a distinct BPS quiver, a different SQM description: all these quivers are related to each other by sequences of elementary mutations, that, at the SQM level, are simply 1d Seiberg--like dualities \cite{CV11,BPSQ}. In the mathematical literature about the categorification of cluster algebras, this statement is known as the fact that mutation equivalent quivers with superpotentials correspond to derived equivalent categories of representations \cite{Keller}.\footnote{ For a more precise statement, the interested reader can consult the very interesting monograph \cite{Keller2}.} The single centered indices satisfy the following generalized--mutation identities: as the BPS quiver undergoes an elementary mutation at a node $\ell$, the single centered indices 
$\Omega_S'$ of the mutated quiver have to satisfy\cite{MPS5,MPSreview}
\be \label{eomtrs}
\Omega_S\left(\sum_i n_i e_i\,; y,t\right) = \Omega_S'\left(\sum_i \mu^S_{\ell}(n_i) \mu_\ell(e_i)\,;y, t\right),
\ee
where $\mu_{\ell}$ is an elementary mutation at $\ell$, that corresponds to the following change of basis:
\be
\mu_\ell(e_i) \equiv \begin{cases} -e_i &\text{if } i = \ell\\ e_i + [\varepsilon B_{i\ell}]_+ e_\ell &\text{if } i \neq\ell\end{cases}\, .
\ee
Here and below, $[x]_+ \equiv \text{max}(0,x)$.
$\mu_\ell(e_i)$ are the basis vectors of the mutated quiver,
while $\mu^S_{\ell}$ is a generalized elementary mutation at $\ell$, defined as follows (see eqn.(3.15) of \cite{MPSreview}, for example)
\be \label{edefmus}
\mu^S_{\ell}(n_i) \equiv \begin{cases}  n_i &\text{for } i\neq \ell \\
n_\ell& \text{for } i=\ell, \, \,  n_j=0\, \, \forall \, j\ne \ell \\
-n_\ell + \sum_{j\neq \ell} n_j [\varepsilon B_{j\ell}]_+ -[\varepsilon \sum_{j\neq \ell} n_j B_{j\ell})]_+&\text{otherwise} .\end{cases}
\ee
$\varepsilon$ takes values $\pm 1$, with the sign depending on whether we are
considering left or right mutations.
Mutation symmetry
implies that for $\gamma\not\parallel e_\ell$, 
$Q(\gamma;p;y,t)$ computed with the original quiver is the same as that
computed with the mutated quiver provided we express $\gamma$ in the new basis
$\mu_\ell(e_i)$,
and choose the $\Omega_S$'s of the new quiver according to \eqref{eomtrs}.
On the other hand for $\gamma\parallel e_\ell$, $Q(\gamma;p;y,t)$ computed with the original quiver\
is equal to $Q(-\gamma;p;y,t)$ computed with the new quiver.

Let us choose $\varepsilon =1$ for definiteness. With this choice we can simplify the 
 $\mu^S_{\ell}(n_\ell)$ formula in the last line of \eqref{edefmus} as:
\be\label{sblam!}
\begin{aligned}
\mu^S_{\ell}(n_\ell) &= -n_\ell + \sum_{j\neq \ell} n_j [B_{j\ell}]_+ + \text{min}\left(0\,,\, -\sum_{j\neq \ell} n_j B_{j\ell}\right)\\
&= -n_\ell + \text{min}\left( \sum_{j\neq \ell} n_j [B_{j\ell}]_+\,,\,\sum_{j\neq \ell} n_j [B_{j\ell}]_+ - \sum_{j\neq \ell} n_j B_{j\ell}\right)\\
&= -n_\ell + \text{min}\left( \sum_{j:B_{j\ell}>0} n_j B_{j\ell}\,,\, \sum_{j: B_{\ell j}>0} n_j B_{\ell j}\right).\\
\end{aligned}
\ee
This can be interpreted as
\be \label{einterpret}
\mu^S_{\ell}(n_\ell) =-n_\ell + \hbox{ min[$\#$ arrows entering node $\ell$, $\#$ arrows leaving node $\ell$]} ,
\ee
where in counting the number of arrows entering (resp. leaving) a node, we have to weight
it with the rank at the node from which it originates (resp. to which it ends). It is easy to see
that even for $\varepsilon=-1$, eq.\eqref{einterpret} and the last line of \eqref{sblam!}
holds.

A simple consequence of the mutation rules \eqref{sblam!}
is that mutation can never take a vector
in $\Gamma^+$ that is not parallel to a basis vector to
a vector that is parallel to a basis vector. To prove this let us assume the
contrary and assume that there is a mutation that gives $\mu_\ell^S(n_k)\ne 0$, 
$\mu_\ell^S(n_i)=0$ for
$i\ne k$. First consider the possibility $\ell=k$. In this case we must have 
$n_i=0$ for $i\ne k$
since $\mu_k^S(n_i)=n_i$. Thus the original vector was parallel to $e_k$ contradicting
our assumption. If on the
other hand $\ell\ne k$, then we must have $n_k=\mu^S_\ell(n_k) \ne 0$, 
$n_i=\mu^S_\ell(n_i)=0$ for $i\ne \ell, k$.
Since there are only two nodes $\ell$ and $k$ where $n_i\ne 0$, there is
only a set of arrows connecting $\ell$ and $k$. Thus using \eqref{einterpret}
we get $\mu_\ell^S(n_\ell)=-n_\ell$. The only way this can vanish is if $n_\ell=0$.
This means that the original vector itself was parallel to 
a basis vector, contradicting our assumption.
This proves the result. Note that this also excludes the case where 
$\mu^S_\ell(n_i)=0$ for all $i$.

If $\mu_\ell^S(n_\ell)$ is negative then the right hand side eq.\eqref{eomtrs} 
vanishes. As a consequence $\Omega_S(\sum n_i e_i\,;\,y,t)$ on the left hand side
must also vanish. This observation will be a key ingredient in our analysis.

\subsection{The property $\omega_0$ and absence of exotics}\label{omegazero}
By the Coulomb branch formula, the absence of exotics conjecture reduces to the statement that the $\Omega_S(\gamma\,;\,y,t)$ are independent of $t$. In fact, we are going to show 
that the much more stronger
\begin{quote}
\noindent{\bf Property $(\omega_0)$}:\emph{ $\Omega_S(\gamma\,;\,y,t) \equiv 0$ for all $\gamma\in\Gamma^+$ not in the basis,}
\end{quote}

\noindent
holds for a set of BPS quivers.
Our strategy to show that $(\omega_0)$ holds true for a given BPS quiver, is to find suitable generalized--mutation sequences that by construction takes outside $\Gamma^+$ any $\gamma$ that is not aligned with a basis vector. 
By the crucial properties in eqns.\eqref{crucial2}--\eqref{eomtrs}, given such a sequence of generalized mutations, $(\omega_0)$ follows. Given that BPS particles can be identified with the objects of a category of quiver representations, once it has been shown that the absence of exotic is satisfied for a given ``master'' theory, the same property has to hold for all those models whose BPS quiver describes a subcategory of the given ``master'' theory category .

\medskip

As a first non--trivial application of the above strategy, in section \ref{SYM} we apply it in the context of pure SYM theories with a simple simply--laced gauge group $G$, thus extending the results of \cite{MMgeom}, and construct some interesting subcategories of these in section \ref{sspecial}. In a forthcoming paper we are going to apply our strategy to other BPS quivers \cite{DZS2}.

\section{The case of pure SYM theories}\label{SYM}

In this section we shall prove property $(\omega_0)$ for the supersymmetric quivers
which appear in computing BPS spectrum of pure ADE gauge theories. Besides proving the
absence of exotics, our result also shows that for these quivers the Coulomb branch formula
gives the result for the spectrum of BPS states in these theories without any further input ({\it e.g.}
knowledge of single centered indices).

\subsection{ Absence of exotics and pure $SU(N)$ SYM}
The BPS quiver of $SU(N+1)$ SYM is the following \cite{BPSQ,CNV,Fiol,cattoy}:
\be
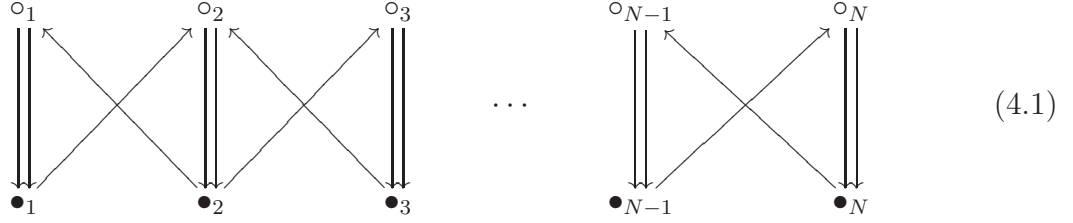
\label{SUN}
\begin{gathered}
\xymatrix{\circ_1\ar@<-0.4ex>[dd]\ar@<0.4ex>[dd]&&\circ_2\ar@<-0.4ex>[dd]\ar@<0.4ex>[dd]&&\circ_3\ar@<-0.4ex>[dd]\ar@<0.4ex>[dd]&&\circ_{N-1}\ar@<-0.4ex>[dd]\ar@<0.4ex>[dd]&&\circ_{N}\ar@<-0.4ex>[dd]\ar@<0.4ex>[dd]\\
&&&&&\cdots\\
\bullet_1\ar[uurr]&&\bullet_2\ar[uull]\ar[uurr]&&\bullet_3\ar[uull]&&\bullet_{N-1}\ar[uurr]&&\bullet_{N}\ar[uull]}\end{gathered}
\ee
Consider $\gamma\in\Gamma^+$ such that
\be
\gamma = \sum_{i=1}^N\left( B_i e_{\bullet_i} + W_i e_{\circ_i}\right).
\ee
Assume that 
\be\label{asSUN}
\sum_{i=1}^N W_i \leq \sum_{i=1}^N B_i\, .
\ee
Mutate over all black nodes of eqn.\eqref{SUN}. This leaves the $W_i$'s unchanged
and maps $B_i \to B_i^\prime$ given by
\bea\label{SUNseq}
B^\prime_1 &= - B_1 + \text{min}(2 W_1, W_2)\\
B^\prime_2 &= - B_2 + \text{min}(2 W_2, W_1+W_3)\\
\vdots\\
B^\prime_{N-1} &= - B_{N-1} + \text{min}(2 W_{N-1}, W_{N-2}+W_N)\\
B^\prime_N &= - B_N + \text{min}(2 W_N, W_{N-1})\, .\\
\eea
Therefore
\be\label{asSUN2}
\sum_{i=1}^N B^\prime_i \leq - \sum_{i=1}^N B_i + 2 \sum_{i=1}^N W_i\leq \sum_{i=1}^N W_i,
\ee
where in the last step we have used \eqref{asSUN}.
Moreover, mutating over all black nodes above, the $SU(N+1)$ quiver is mapped into itself up to exchanging the black with the white nodes. Now mutating all white nodes above, 
one obtains the same result with $B_i\leftrightarrow W_i$ and $W_i\leftrightarrow B^\prime_i$. Eventually, one lands on the inequality:
\be \label{enext}
\sum_{i=1}^N W^\prime_i \leq - \sum_{i=1}^N W_i + 2 \sum_{i=1}^N B^\prime_i \leq \sum_{i=1}^N B^\prime_i.
\ee
where in the last step we have used \eqref{asSUN2}.
Iterating this process either one reaches a fixed point where 
$\sum_i B_i=\sum_i W_i$, or the sum turns negative. 

Let us examine the conditions for a fixed point. 
We already know from the analysis below \eqref{einterpret} that the fixed point must have
at least two non-zero elements in the set  $\{W_i, B_i\}$.
A necessary condition for the fixed point is that the following inequalities are satisfied:
\bea \label{efeqa}
&2 W_1\leq W_2\\
&2 W_2\leq W_1+W_3\\
&\vdots\\
&2 W_{N-1}\leq W_{N-2}+W_N\\
&2 W_N\leq W_{N-1}.\\
\eea
There are also similar conditions involving the $B_i$'s. 
Finally we must have $B_i=W_i$ for each $i$ to ensure that individual
$B_i$'s and $W_i$'s cannot be decreased by repeated application of
the mutation sequence, eventually turning one of them negative.
We now observe that
the conditions \eqref{efeqa} are simply $C_{A_{N}} \cdot \vec{W} \leq 0$,
where $C_{A_N}$ is the Cartan matrix of $A_N$ and we have collected together the $W_i$'s in a single vector: $\vec{W}=(W_1,\dots,W_N)$. Since the ADE Cartan matricies are positive definite, these equations cannot have any solution with 
$W_i\ge 0$, not all $W_i$ zero.
This leads us to the conclusion that mutation eventually turns either $\sum_i W_i$
or $\sum_i B_i$ negative, taking us outside $\Gamma^+$. 
Hence $\Omega_S(\gamma;y,t)$ must
vanish.

If, instead,
\be
\sum_{i=1}^N B_i \leq \sum_{i=1}^N W_i
\ee
we can start the mutation sequence by mutating on all the white nodes first. Exactly the same argument with black nodes and white nodes exchanged applies. This shows that $(\omega_0)$ holds for $SU(N+1)$ SYM, and the absence of exotics follows.

\subsection{Absence of exotics and pure $SO(2N)$ SYM}
The BPS quiver for $SO(2N)$ SYM is the following \cite{CNV,BPSQ,cattoy}:
\be
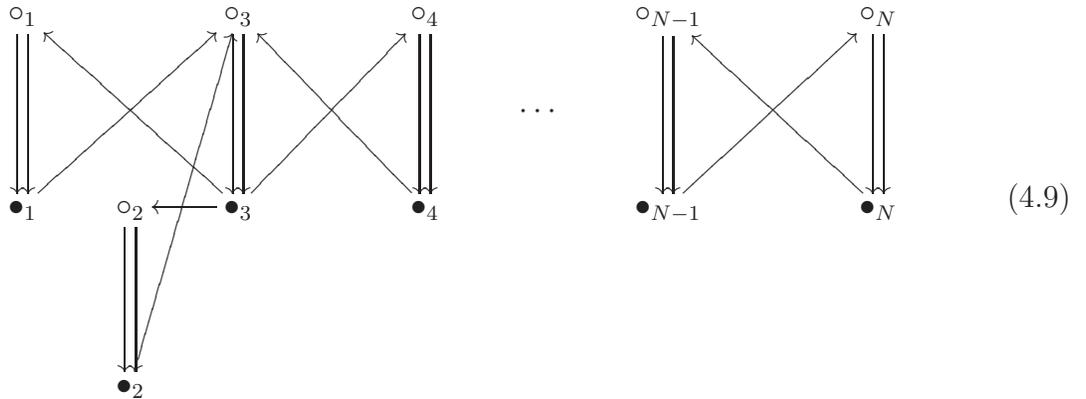
\begin{gathered}
\xymatrix{\circ_1\ar@<-0.4ex>[dd]\ar@<0.4ex>[dd]&&\circ_3\ar@<-0.4ex>[dd]\ar@<0.4ex>[dd]&&\circ_4\ar@<-0.4ex>[dd]\ar@<0.4ex>[dd]&&\circ_{N-1}\ar@<-0.4ex>[dd]\ar@<0.4ex>[dd]&&\circ_{N}\ar@<-0.4ex>[dd]\ar@<0.4ex>[dd]\\
&&&&&\cdots\\
\bullet_1\ar[uurr]& \circ_2\ar@<-0.4ex>[dd]\ar@<0.4ex>[dd] &\bullet_3 \ar[l] \ar[uull]\ar[uurr]&&\bullet_4\ar[uull]&&\bullet_{N-1}\ar[uurr]&&\bullet_{N}\ar[uull]\\
&&&&&\\
&\bullet_2\ar[uuuur] &&&&\\
}
\end{gathered}
\ee
Assuming a condition equivalent to eqn.\eqref{asSUN}, we first mutate on all the black nodes as before. Again the quiver gets mapped onto itself up to exchange of the black and the white nodes. By eqn.\eqref{sblam!}, since the black nodes are always sinks with respect to the vertical Kroneckers and sources with respect to the other 
white nodes, we are able to reproduce exactly the same type of chains of inequalities 
\eqref{asSUN2}, \eqref{enext} we encountered in the $SU(N)$ case. Only the condition for the fixed point is different and now reads:
\bea \label{ednfixed}
&2 W_1 \leq W_3\\
&2W_2 \leq W_3\\
&2 W_3 \leq W_1 + W_2 + W_4\\
&2 W_4 \leq W_3 + W_5\\
\vdots\\
&2 W_{N-1}\leq W_{N-2} + W_N\\
&2 W_N \leq W_{N-1} ,
\eea
and $B_i=W_i$ for each $i$.
Eq.\eqref{ednfixed} can be expressed as 
\be
C_{D_N} \cdot \vec{W} \leq 0,
\ee
where $C_{D_N}$ is the Cartan matrix of $D_N$ type. By positive definiteness of ADE Cartan matrices, this system has no solution with $W_i\ge 0$ and at least one $W_i$
non-zero. Thus we 
conclude that the series of mutations will take us outside $\Gamma^+$, giving 
$\Omega_S(\gamma;y,t)=0$.

\begin{figure}
\begin{center}

$$
E_6 \qquad \colon \quad
\begin{gathered}
\xymatrix{\circ_1\ar@<-0.4ex>[dd]\ar@<0.4ex>[dd]&&\circ_2\ar@<-0.4ex>[dd]\ar@<0.4ex>[dd]&&\circ_3\ar@<-0.4ex>[dd]\ar@<0.4ex>[dd]&&\circ_5\ar@<-0.4ex>[dd]\ar@<0.4ex>[dd]&&\circ_{6}\ar@<-0.4ex>[dd]\ar@<0.4ex>[dd]\\
&&&&&\\
\bullet_1\ar[uurr]&&\bullet_2\ar[uurr]\ar[uull]& \circ_4\ar@<-0.4ex>[dd]\ar@<0.4ex>[dd] &\bullet_3 \ar[l] \ar[uull]\ar[uurr]&&\bullet_5\ar[uull]\ar[uurr]&&\bullet_{6}\ar[uull]\\
&&&&&\\
&&&\bullet_4\ar[uuuur] &&&&\\
}
\end{gathered}
$$

\bigskip

$$
E_7 \qquad \colon \quad
\begin{gathered}
\xymatrix{\circ_1\ar@<-0.4ex>[dd]\ar@<0.4ex>[dd]&&\circ_2\ar@<-0.4ex>[dd]\ar@<0.4ex>[dd]&&\circ_3\ar@<-0.4ex>[dd]\ar@<0.4ex>[dd]&&\circ_5\ar@<-0.4ex>[dd]\ar@<0.4ex>[dd]&&\circ_{6}\ar@<-0.4ex>[dd]\ar@<0.4ex>[dd]&&\circ_{7}\ar@<-0.4ex>[dd]\ar@<0.4ex>[dd]\\\
&&&&&\\
\bullet_1\ar[uurr]&&\bullet_2\ar[uurr]\ar[uull]& \circ_4\ar@<-0.4ex>[dd]\ar@<0.4ex>[dd] &\bullet_3 \ar[l] \ar[uull]\ar[uurr]&&\bullet_5\ar[uull]\ar[uurr]&&\bullet_6\ar[uull]\ar[uurr]&&\bullet_{7}\ar[uull]\\
&&&&&\\
&&&\bullet_4\ar[uuuur] &&&&\\
}
\end{gathered}
$$

\bigskip

$$
E_8 \qquad \colon \quad
\begin{gathered}
\xymatrix{\circ_1\ar@<-0.4ex>[dd]\ar@<0.4ex>[dd]&&\circ_2\ar@<-0.4ex>[dd]\ar@<0.4ex>[dd]&&\circ_3\ar@<-0.4ex>[dd]\ar@<0.4ex>[dd]&&\circ_5\ar@<-0.4ex>[dd]\ar@<0.4ex>[dd]&&\circ_{6}\ar@<-0.4ex>[dd]\ar@<0.4ex>[dd]&&\circ_{7}\ar@<-0.4ex>[dd]\ar@<0.4ex>[dd] &&\circ_{8}\ar@<-0.4ex>[dd]\ar@<0.4ex>[dd]\\\
&&&&&\\
\bullet_1\ar[uurr]&&\bullet_2\ar[uurr]\ar[uull]& \circ_4\ar@<-0.4ex>[dd]\ar@<0.4ex>[dd] &\bullet_3 \ar[l] \ar[uull]\ar[uurr]&&\bullet_5\ar[uull]\ar[uurr]&&\bullet_6\ar[uull]\ar[uurr]&&\bullet_7\ar[uull]\ar[uurr]&&\bullet_{8}\ar[uull]\\
&&&&&\\
&&&\bullet_4\ar[uuuur] &&&&\\
}
\end{gathered}
$$

\caption{Quivers for SYM with exceptional simply-laced Lie groups \cite{CNV,BPSQ,cattoy}.}\label{exceptionals}
\end{center}
\end{figure}
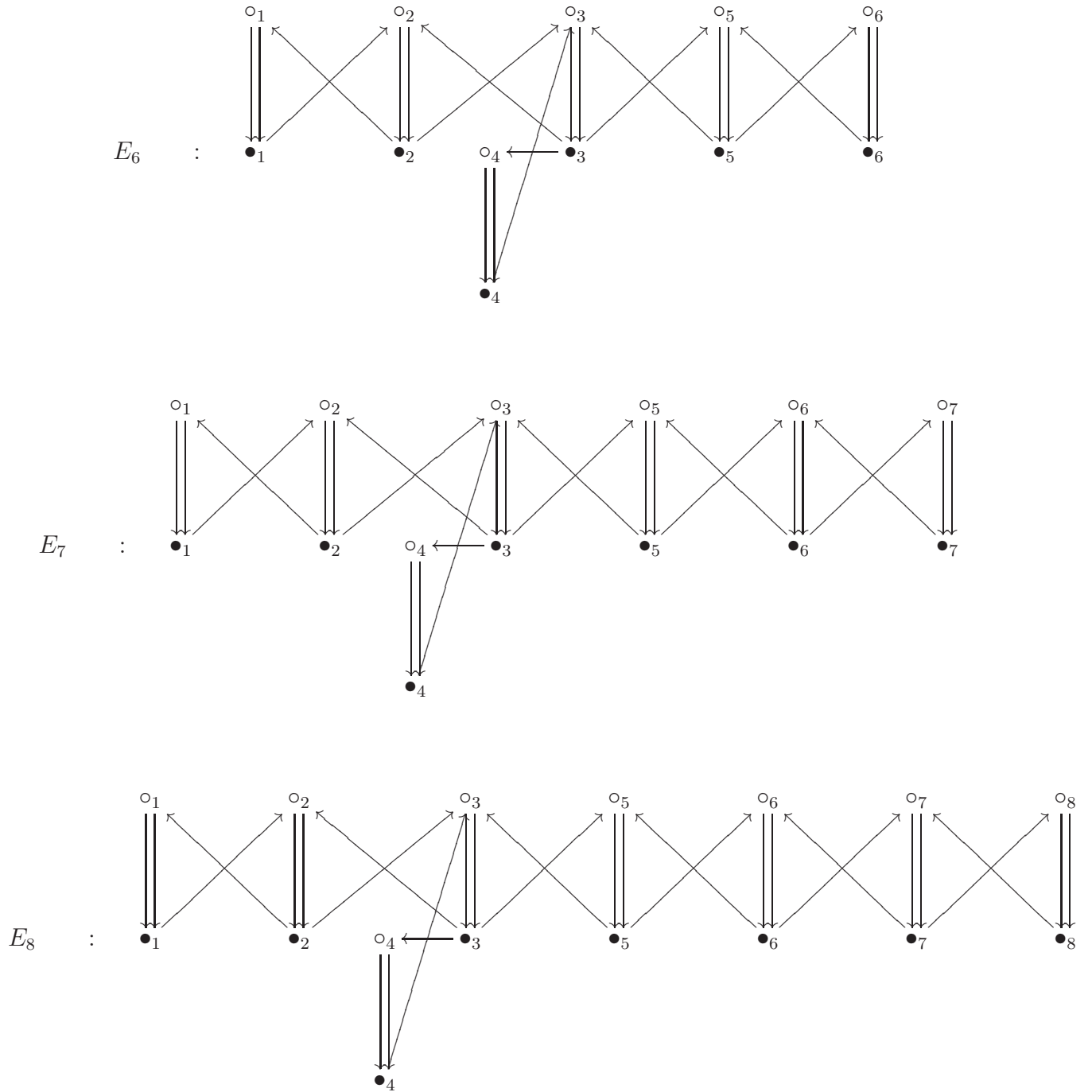

\subsection{Absence of exotics and pure $E_{6,7,8}$ SYM}
The BPS quivers of exceptional simple simply laced SYM theories are drawn in figure \ref{exceptionals}. Again repeating words for words the analysis of the previous cases one obtains
the fixed point condition
\be
C_{G} \cdot \vec{W} \leq 0, \qquad\text{with } G=E_6,E_7,E_8, \qquad \vec B = \vec W,
\ee
as a condition for a fixed point along the decreasing chain of inequalities implied by our formula \eqref{sblam!}. As previously, this cannot be satisfied by a non-trivial $\vec{W}$. 
This concludes our argument showing the absence of exotics from pure SYM theories with simple simply--laced gauge groups. 

\section{Specialization: from SYM to $G[\ci]$ theories} \label{sspecial}
An interesting consequence of the fact that the SYM quivers have no exotic BPS states is the following. As we have discussed in section \S.\ref{squiver} the absence of exotics is a property of the category of representations of the SYM quivers. As such, it has to be shared by all proper subcategories. By the specialization argument of \cite{cattoy,nsl} many known models have quivers whose categories of representations are embedded canonically into the ones of the SYM quivers as controlled subcategories. As a corollary of the property $\omega_0$ of the pure SYM theories, it follows that all these models obey the no exotics conjecture.

The more trivial instance of specialization involves the specialization at an external node of a given Dynkin graph. At the quiver level this operation is essentially the following:

\be\label{spec1}
\begin{gathered}
\xymatrix{
\circ\ar@{<-}[ddrr]&&\ar@{<-}[ddll]\circ\ar@{<-}[ddrr]&&\cdots\ar@{<-}[ddll]\\ 
&&\\ 
\bullet\ar@<-0.4ex>@{<-}[uu]\ar@<0.4ex>@{<-}[uu]&&\ar@<-0.4ex>@{<-}[uu]\ar@<0.4ex>@{<-}[uu]\bullet&&\cdots
}
\end{gathered}
\qquad\longrightarrow\qquad
\begin{gathered}
\xymatrix{
&\ar@{<-}[dl]\circ\ar@{<-}[ddrr]&&\cdots\ar@{<-}[ddll]\\ 
\ast\ar@{<-}[dr]&&\\ 
&\ar@<-0.4ex>@{<-}[uu]\ar@<0.4ex>@{<-}[uu]\bullet&&\cdots
}
\end{gathered}
\ee
or, for external nodes at a trivalent vertex:
\be\label{spec1b}
\begin{gathered}
\begin{gathered}
\xymatrix{\circ\ar@<-0.4ex>[dd]\ar@<0.4ex>[dd]&&\circ\ar@<-0.4ex>[dd]\ar@<0.4ex>[dd]&&\circ\ar@<-0.4ex>[dd]\ar@<0.4ex>[dd]\\
&&&&&\cdots\\
\bullet\ar[uurr]& \circ\ar@<-0.4ex>[dd]\ar@<0.4ex>[dd] &\bullet \ar[l] \ar[uull]\ar[uurr]&&\bullet\ar[uull]\\
\cdots&&&&&\\
&\bullet\ar[uuuur] &&\\
}
\end{gathered}\\
\xymatrix{\ar[d]\\
\\}\\
\begin{gathered}
\xymatrix{&&\circ\ar@<-0.4ex>[dd]\ar@<0.4ex>[dd]&&\circ\ar@<-0.4ex>[dd]\ar@<0.4ex>[dd]\\
\ast\ar[urr]&&&&&\cdots\\
& \circ\ar@<-0.4ex>[dd]\ar@<0.4ex>[dd] &\bullet \ar[l] \ar[ull]\ar[uurr]&&\bullet\ar[uull]\\
\cdots&&&&&\\
&\bullet\ar[uuuur] &&\\
}
\end{gathered}
\end{gathered}
\ee

\noindent accompanied by a corresponding specialization of the superpotential \cite{nsl}. At the level of gauge theory, this is a map that goes from $G$ to $U(1) \times H \subset G$, and promotes $U(1)$ to a flavor symmetry by sending the corresponding coupling to zero.\footnote{ This decoupling limit is realized by means of a light subcategory at the level of representation theory \cite{ cattoy, alf, nsl}. The specialized quivers are effective quiver descriptions of the corresponding light categories.} The matter content is determined via
\be
\text{Adj}_G \to \text{Adj}_H \oplus R_H,
\ee
where $R_H$ is a representation of $H$ determined by $G$. All the systems that can be obtained from simply--laced $G$ by a specialization at an external node of the corresponding Dynkin graph are summarized in table \ref{tsp}\cite{cattoy}.

\begin{table}
$$
\begin{tabular}{ccc}
$G$&$H$ & $R$\\
\hline
$SU(N+1)$&$SU(N)$ & $\mathbf{N}$\\
$SO(2N+2)$&$SU(N)$ & $\mathbf{N(N-1)/2}$\\
$E_6$&$SU(6)$ & $\mathbf{20}$\\
$E_7$&$SU(7)$ & $\mathbf{35}$\\
$E_8$&$SU(8)$ & $\mathbf{56}$\\
\hline
$SO(2N+2)$&$SO(2N)$ & $\mathbf{2N}$\\
$E_6$&$SO(10)$ & $\mathbf{16}$\\
$E_7$&$SO(12)$ & $\mathbf{32}$\\
$E_8$&$SO(14)$ & $\mathbf{64}$\\
\hline
$E_7$&$E_6$ & $\mathbf{27}$\\
$E_8$&$E_7$ & $\mathbf{56}$\\
\hline
\end{tabular}
$$
\caption{All systems obtained from a simply--laced group $G$ by a single specialization of the type in eqn.\eqref{spec1} \cite{cattoy}.}\label{tsp}
\end{table}

Analogously, one can study the specialization at an internal node of the Dynkin graph $G$. For example consider the case in which no trivalent vertexes are involved (again we write down only the effect on the quiver):

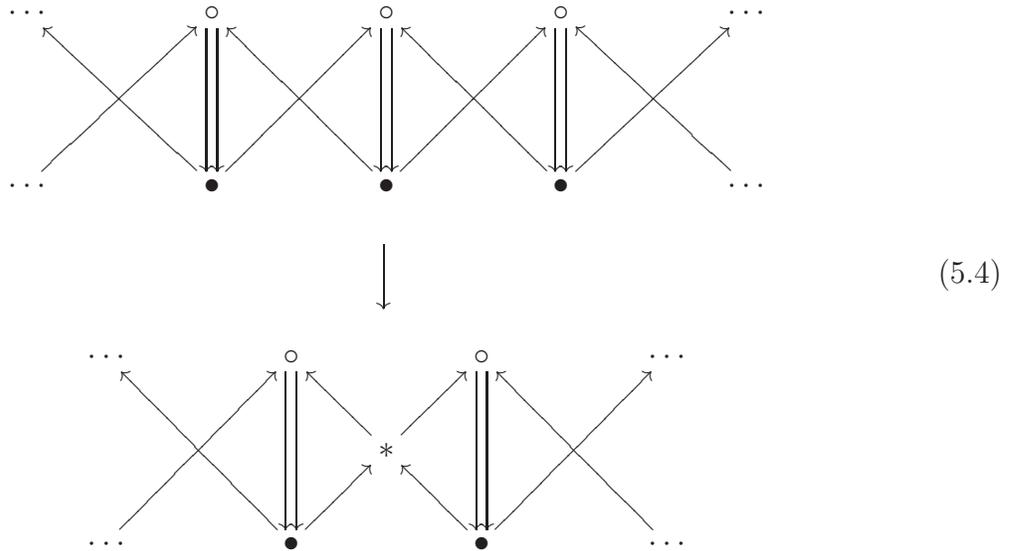
\be
\begin{gathered}
\begin{gathered}
\xymatrix{
\cdots&&\circ\ar@<-0.4ex>[dd]\ar@<0.4ex>[dd]&&\circ\ar@<-0.4ex>[dd]\ar@<0.4ex>[dd]&&\circ\ar@<-0.4ex>[dd]\ar@<0.4ex>[dd]&&\cdots\\
\\
\cdots\ar[uurr]&&\bullet\ar[uurr]\ar[uull]&&\bullet\ar[uurr]\ar[uull]&&\bullet\ar[uurr]\ar[uull]&&\cdots\ar[uull]\\
}
\end{gathered}\\
\xymatrix{\ar[d]\\
\\}\\
\begin{gathered}
\xymatrix{
\cdots&&\circ\ar@<-0.4ex>[dd]\ar@<0.4ex>[dd]& &\circ\ar@<-0.4ex>[dd]\ar@<0.4ex>[dd]&&\cdots\\
& &&\ast\ar[ur]\ar[ul]\\
\cdots\ar[uurr]&&\bullet\ar[ur]\ar[uull]& &\bullet\ar[uurr]\ar[ul]&&\cdots\ar[uull]\\
}
\end{gathered}
\end{gathered}
\ee

An example of this type of specialization is the following breaking:
\bea
SU(N+1) &\rightarrow SU(N-k+1)\times SU(k)\\
\text{Adj}_{SU(N+1)}&\rightarrow\text{Adj}_{SU(N-k+1)}\oplus (\mathbf{N-k},\mathbf{k-1})\oplus \text{Adj}_{SU(k)}.
\eea
More generally, consider a simply--laced Dynkin graph $G$, of rank $r$. 
Choose a subset $\ci$ of the set of nodes of $G$ such that each pair of elements of $\ci$ is not connected by an edge of $G$. To each element of $\ci$ corresponds an `elementary' specialization of the type we discussed above. Removing from $G$ the set of nodes $\ci$ one is left with the Dynkin graph of a subgroup of $G$ of the form $H_1 \times H_2 \times\dots \times H_k$. The various groups $H_i$ are, of course, all simply--laced. The matter of the corresponding lagrangian $\cn=2$ system is determined by the decomposition of the adjoint representation of $G$ along the breaking pattern $G \rightarrow H_1\times H_2 \times\dots\times H_k$. Following \cite{cattoy} these models are conveniently labeled $G[\ci]$. All quivers of the $G[\ci]$ models are quivers of a suitable (light) subcategories of the canonical representative of the mutation class of the $G$ SYM quiver. By our result all these systems are automatically free of exotics. A natural question is whether one can show that $(\omega_0)$ is valid for these systems too. A partial answer can be found in appendix \ref{omegazz}.

\section*{Acknowledgements}
We thank Sergio Cecotti, Clay Cordova, Jan Manschot, Gregory Moore, Boris Pioline, and Cumrun Vafa for useful discussions.
The work of MDZ is supported by the NSF grant PHY-1067976.
The work of A.S. was
supported in part by the 
DAE project 12-R\&D-HRI-5.02-0303 and J. C. Bose fellowship of 
the Department of Science and Technology, India.

\appendix

\section{Appearance of exotic states upon crossing a wall of the third kind} \label{sa}

In this appendix we shall describe a specific example in which the spectrum of BPS
states with no exotics acquire exotic states after crossing a wall of the third kind.

Let us suppose that in some region of the moduli space the BPS spectrum consists of
a short multiplet tensored with (1/2,0) representation. This is non-exotic. The 
$SU(2)_{\text{spin}}\times SU(2)_R$ representation content of this is
\be \label{ethree}
(1/2, 1/2) + (1,0) + (0,0)\, .
\ee
Now suppose that we encounter a wall of the third kind on which a long multiplet
tensored with representation (0,1) accidentally has BPS saturated mass. The 
representation content of this state is 
\be \label{efour}
(1,1) + (0,1) + (0,2) + (0,1) + (0,0) + (0,1) + (1/2, 3/2) + (1/2, 1/2) + (1/2, 3/2) + (1/2, 1/2) \, .
\ee
Now we can see that (\ref{ethree}) and (\ref{efour}) together contains all the states
required to build a long multiplet tensored with (0,0) representation, as given in 
eq.(\ref{etwo}). So as we cross the wall of the third kind to the other side, this
subset could become massive, forming a genuine long multiplet.  We are then left with
BPS states in the representation:
\be \label{efive}
(1,1) +  (0,2) + (0,1) + (0,1) + (1/2, 3/2) + (1/2, 3/2) + (1/2, 1/2) \, .
\ee
It is easy to convince oneself that there is no way any subset of these can combine
into a long multiplet. So the spectrum is stable, till we encounter another wall of the
third kind.  It is also easy to see that the spectrum can be interpreted as a short
multiplet tensored with
\be \label{esix}
(0,3/2) + (1/2, 1)
\ee
representation. This certainly contains exotics.

\section{BPS quivers and absence of exotics from class $\cs[A_1]$}\label{ovvieta}
In this appendix we discuss a different argument that proves the no exotics for the theories of class $\cs[A_1]$ with BPS quivers: for these models, it is an obvious corollary of the Geiss--Labardini-Fragoso--Shr\"oer theorem on the representation type of Jacobian algebras\cite{superpowers}.

By standard arguments in geometric invariant theory \cite{Mumford,HKLR}, the computation of $\cm(\gamma,p)$ from the quiver SQM can be rephrased entirely in terms of the representation theory of a quiver with superpotential \cite{DWZ1}, the BPS quiver of the model in question \cite{CV11,BPSQ}. We shall now describe how
the absence of exotics can be proved directly from representation theory for theories of class $\cs[A_1]$ \cite{Gaiotto,GMN_Hitchin} with BPS quivers.  These models are obtained by compactification of the $A_1$ $(2,0)$ theory on a Riemann surface with genus $g$, $p$ regular punctures, at the locations of quadratic p\^oles of the Hitchin field, and $b$ irregular punctures located where the Hitchin field has p\^oles of order $c_s +2$, $c_s\geq1$, $s=1,...,b$, the so called UV curve. The vast majority of the models of class $\cs[A_1]$ has the quiver property \cite{CV11,BPSQ2}: the BPS quivers are precisely the quivers with superpotential that arise from ideal triangulations of the UV surface \cite{FST,FT}. More details about the physics associated to these algebras can be found in \S.4 of \cite{cattoy}. Necessary and sufficent conditions for having an ideal triangulations are as follows: For $g=0$, if $b=0$ one needs $p\geq 3$, if $p=0,1,2$, $b\geq1$ ; for $g\geq 1$, at least one in between $p$ and $b$ has to be $\geq 1$. As a consequence of the Geiss--Labardini-Fragoso--Shr\"oer theorem on the representation type of Jacobian algebras\cite{superpowers}, all the Jacobian algebras corresponding to ideal triangulations are tame. This means that their irreducible representations are either rigid or come in $\mathbb{P}^1$ moduli. Since stable representations are in particular irreducible, this entails that the BPS spectra of complete models contains only hypers and vectors, and, in particular, no exotics states. It would be very interesting to understand whether the stronger property $(\omega_0)$ extends to this class of quivers as well.

\section{Does $(\omega_0)$ extend to all $G[\ci]$ models?}\label{omegazz}

In some cases it is possible to 
 use the techniques of \S\ref{SYM}
to
directly prove the vanishing of $\Omega_S(\gamma\,;\,y,t)$ for the quivers
described in \S\ref{sspecial} when
$\gamma$ is not a basis vector. Take for example the
quiver shown in \eqref{spec1}. Assuming that 
$\sum_i W_i \le \sum_i B_i$, we proceed by first mutating all the black nodes.
This leads to eq.\eqref{asSUN2}. Next we mutate all the white nodes, leading to
\eqref{enext}. Proceeding this way we see that eventually either $\sum_i B_i$ or
$\sum_i W_i$ turns negative, or we reach a fixed point. The fixed point condition 
remains almost the same except that the first equation in \eqref{efeqa}  is replaced
by $2 W_1 \le W_2+N_*$ where $N_*$ is the rank of the node labelled by $*$.
There is also a further condition $N_* \le W_1$ (and also $N_*\le B_1$) which is needed
to ensure that we cannot turn the rank of $*$ negative by mutating on $*$. These
equations may be expressed as
\be \label{efeq}
\begin{pmatrix}
1 & -1 & 0 & 0 & \cdots \cr
-1 & 2 & -1 & 0 & \cdots\cr
0 & -1 & 2 & -1 & \cdots\cr
\cdot & \cdot & \cdot & \cdot & \cdots \cr
\cdot & \cdot & \cdot & \cdot & \cdots
\end{pmatrix}
\begin{pmatrix}
N_* \cr W_1 \cr W_2\cr \cdot \cr W_N
\end{pmatrix} \le 0\, ,
\ee
where $\le 0$ condition needs to be satisfied by each entry of the resulting vector.
Using a recursive method it is easy to show that these equations have no 
non-trivial solution 
with non-negative $N_*$ and $W_i$. For this we can add the first equation to the
second equation to write another inequality
\be 
\begin{pmatrix}
1 & -1 & 0 & 0 & \cdots \cr
0 & 1 & -1 & 0 & \cdots\cr
0 & -1 & 2 & -1 & \cdots\cr
\cdot & \cdot & \cdot & \cdot & \cdots \cr
\cdot & \cdot & \cdot & \cdot & \cdots
\end{pmatrix}
\begin{pmatrix}
N_* \cr W_1 \cr W_2\cr \cdot \cr W_N
\end{pmatrix} \le 0\, .
\ee
We now notice that second to the last equation has the same structure as 
eq.\eqref{efeq} with $N\to N-1$ and the replacement
\be
N_*\to W_1, \quad W_1\to W_2, \quad \cdots , \quad W_{N-1}\to W_N\, .
\ee
Thus a necessary condition
for \eqref{efeq} to hold is that a similar inequality with $N\to N-1$ 
should also have non-trivial solution. 
By repeating this analysis we can conclude that a
necessary condition for \eqref{efeq} to hold is that a similar inequality with $N=1$
must have non-trivial solution. 
Since the corresponding matrix $\begin{pmatrix} 1 & -1 \cr -1 & 2 \end{pmatrix}$
is positive definite, corresponding inequality  has no solution. Thus
the original inequality also has no solution.

In each of the other cases, we can use repeated mutation of the black nodes followed
by that of white nodes to show that we either  make one or more
of the ranks negative leading to vanishing of the corresponding $\Omega_S$ or
reach a fixed point. In fact since the left and right mutations, which are inverses
of each other, have the same action on $\Omega_S$, we cannot really reach a fixed
point starting from a configuration that is not a fixed point; we have to be at the fixed
point from the beginning. The second step, showing that no fixed point exists, does 
not however go through for all the quivers. Nevertheless this analysis gives us 
constraints on the $B_i$'s and $W_i$'s for which $\Omega_S$ can be non-zero. 
Presumably by further analysis, {\it e.g.} mutating on the $*$ nodes, we can show that
these $\Omega_S$'s also vanish, but we have not carried out a detailed analysis for all
the cases. This is a very interesting question we are going to address \cite{DZS2}.\footnote{ A comment for {\it cognoscendi}: for the moment, using Coxeter--factorized sequences of mutations \cite{Arnold1}, we are able to construct $\Omega_S$--decreasing sequences that much likely do not admit fixed points as well. We have decided to discuss these result in a followup paper, to keep technicalities at minimum in this note.}

\end{document}